\documentclass[runningheads]{llncs}
\usepackage[T1]{fontenc}
\usepackage{graphicx}
\usepackage{amsfonts}
\usepackage{amssymb}
\usepackage{amsmath}

\begin{document}
\title{Masked Conditional Diffusion Models for Image Analysis with Application to Radiographic Diagnosis of Infant Abuse} 
\titlerunning{Masked Conditional Diffusion Models for Image Analysis}
% If the paper title is too long for the running head, you can set
% an abbreviated paper title here
%
\author{Shaoju Wu \and
Sila Kurugol \and
Andy Tsai}
\authorrunning{S. Wu et al.}
% First names are abbreviated in the running head.
% If there are more than two authors, 'et al.' is used.
%
%\institute{Institute withheld}
\institute{Boston Children's Hospital and Harvard Medical School, Boston, MA, USA \email{shaoju.wu@childrens.harvard.edu}}
\maketitle              % typeset the header of the contribution
\begin{abstract}
The classic metaphyseal lesion (CML) is a distinct injury that is highly specific for infant abuse. It commonly occurs in the distal tibia. To aid radiologists detect these subtle fractures, we need to develop a  model that can flag abnormal distal tibial radiographs (i.e. those with CMLs). Unfortunately, the development of such a model requires a large and diverse training database, which is often not available. To address this limitation, we propose a novel generative model for data augmentation. Unlike previous models that fail to generate data that span the diverse radiographic appearance of the distal tibial CML, our proposed masked conditional diffusion model (MaC-DM) not only generates realistic-appearing and wide-ranging synthetic images of the distal tibial radiographs with and without CMLs, it also generates their associated segmentation labels. To achieve these tasks, MaC-DM combines the weighted segmentation masks of the tibias and the CML fracture sites as additional conditions for classifier guidance. The augmented images from our model improved the  performances of ResNet-34 in classifying normal radiographs and those with CMLs. Further, the augmented images and their associated segmentation masks enhanced the performance of the U-Net in labeling areas of the CMLs on distal tibial radiographs.

\end{abstract}

\section{Introduction}
Child abuse constitutes a major public health problem, with 9.2/1000 children abused per year \cite{ChildMaltreatment}. Infants ($\le$1-year-old) are particularly vulnerable to abuse, and constitute 42\% of children who die from inflicted injuries \cite{ChildMaltreatment}. One highly specific fracture for infant abuse is the classic metaphyseal lesion (CML) \cite{coley2013caffey,flaherty2014evaluating,kleinman2011prevalence,servaes2016etiology}. The CML is a unique injury characterized by a fracture plane that courses along the long bone metaphysis \cite{kleinman1986metaphyseal,kleinman1995relationship,kleinman1995inflicted,tsai2014high}. These CMLs can occur in any of the long bones, but the distal tibia is one of the most common sites. CML poses a diagnostic challenge to radiologists due to its variable and frequently subtle radiographic appearances. We need to develop a computer algorithm that can automatically flag potentially abnormal radiographs (i.e. those with CML fractures) for special attention during customary radiologic assessment. Unfortunately, the development of such a model requires a large and diverse training database, which is not available even at large pediatric centers. To address this limitation, a generative model for data augmentation is required.

To generate synthetic training data, we typically employ traditional deep learning generative models such as generative adversarial networks (GANs) \cite{goodfellow2020generative,han2021madgan,kearney2020attention}. However, these models, such as CycleGAN \cite{zhu2017unpaired}, can be difficult to train and may generate low-quality and repetitive images when trained with limited datasets \cite{saxena2021generative}. Recently, a new generative model called denoising diffusion probabilistic model (DDPM) was shown to be more effective in generating realistic synthetic images and has a more straightforward training process \cite{dhariwal2021diffusion}. It has demonstrated utility in data augmentation for brain tumor detection \cite{wolleb2022diffusion}, MR image denoising \cite{peng2022towards}, and image generation \cite{kim2022diffusion}. One major advantage of DDPM is its ability to incorporate conditional information, including class labels, to generate high quality conditional images \cite{dhariwal2021diffusion}. However, current DDPM-based methods \cite{kim2022diffusion,peng2022towards,wolleb2022diffusion} are not applicable for our particular problem of CML fracture detection. Inspired by the conditional DDPM that utilizes conditional mask images as additional inputs for image inpainting \cite{rombach2022high}, we propose a novel mask conditional diffusion model (MaC-DM) that incorporates weighted segmentation masks of the tibias and the CML fractures as priors in generating realistic and diverse synthetic radiographic images of distal tibia with and without CMLs.

We aim to perform binary classification to categorize distal tibial radiographs as either normal or abnormal. To improve the performance of the baseline classification model trained with limited clinical data, we used our proposed MaC-DM for data augmentation. To highlight the added value of this methodology as a data augmenter, we compared our methodology to other common generative models. In addition to generating synthetic images, our proposed model produces masks of the CML fracture sites, which can be used as additional augmented dataset to train a segmentation algorithm in labeling areas of the CMLs in the distal tibial radiographs. In essence, our contributions are three-folds. One, we developed a new generative diffusion model which utilizes both class labels and segmentation masks as conditions to generate realistic and diverse normal and abnormal radiographic images. Two, the segmentation mask obtained from our generative model can be used to improve the segmentation of the CML fracture region, which is beneficial for verification of automated classification. Three, this is the first application of generative diffusion models to augment a CML database to improve the diagnosis of infant abuse. 

\section{Methods}   
In this paper, we propose a new generative model for data augmentation of radiographic images called MaC-DM. This model consists of two stages: 1) training the denoising network, and 2) sampling or generation of new data by integrating the conditional information (Fig. 1). In the first stage, we train two different diffusion models using radiographic images, their associated conditional information in the form of their binary class labels (CML fracture versus normal), and their segmentation masks (for both the distal tibia and the CML fracture region). We train one diffusion model to classify noisy CML and normal radiographs, which we use for classifier guidance during sampling; and train another diffusion model to remove noise from the noisy images, which we compute after adding noise to the training dataset. In the second stage, we use the trained diffusion model with classifier guidance to generate 1) synthetic radiographs of the distal tibia, and 2) their corresponding tibial and CML segmentation masks.

\subsection{Diffusion model}                             
DDPM is a special type of generative model designed to transform a Gaussian distribution to an empirical one \cite{ho2020denoising,nichol2021improved}. It is embodied by a forward and a reverse process. In particular, given a data sample $x_{0} \sim q(x)$, the forward process gradually adds small amounts of noise to the input data from time steps $t=0$ to $T$, according to a pre-determined noise schedule, as described by the following:
\begin{equation}
q(x_{t}|x_{t-1})=\mathcal{N}(x_{t};\sqrt{1-\beta_{t}}x_{t-1},\beta_{t}\mathbf I) 
\end{equation}
where $x_{t}$ and $\beta_{t}$ represent the noisy image and the noise variance at time step $t$, respectively; with the noisy image $x_{t}$ described by
$x_{t}=\sqrt{\bar{\alpha}_{t}}x_{0}+\sqrt{1-\bar{\alpha}_{t}}\epsilon
$
where ${\alpha}_{t}=1-\beta_{t}$ and $\bar{\alpha}_{t}=\prod_{t=1}^{T}\alpha_{t}$. In contrast, in the reverse process, we generate samples of the denoised image as follows:
 \begin{equation}
p_\theta(\mathbf{x}_{t-1} \vert \mathbf{x}_t)=\mathcal{N}(\mathbf{x}_{t-1}; {\mu}_\theta(\mathbf{x}_t, t),\sigma^{2}_{t}\mathbf I)
\end{equation}
where ${\mu}_\theta(\mathbf{x}_t, t)$ is the mean value learned by DDPM, and $\sigma^{2}_{t}\mathbf I$ is the fixed variance.

Typically, the sampling process of DDPM is slow due to the large number of time steps required for denoising. To speedup this process, we use the denoising diffusion implicit model (DDIM) for sampling acceleration \cite{song2020denoising}. To train our model for denoising, we employ the following loss functional ~\cite{ho2020denoising}:
\begin{equation}
L= \mathbb{E}_{t \sim [1,T], x_{0}, \epsilon_t}[\| \epsilon-\epsilon_\theta(x_t, t)\|_{2}^2]  
\end{equation}
where $\epsilon \sim \mathcal{N}(0,I)$, and $\epsilon_\theta$ is the learned diffusion model.

\begin{figure}[t]
\includegraphics[width=\textwidth]{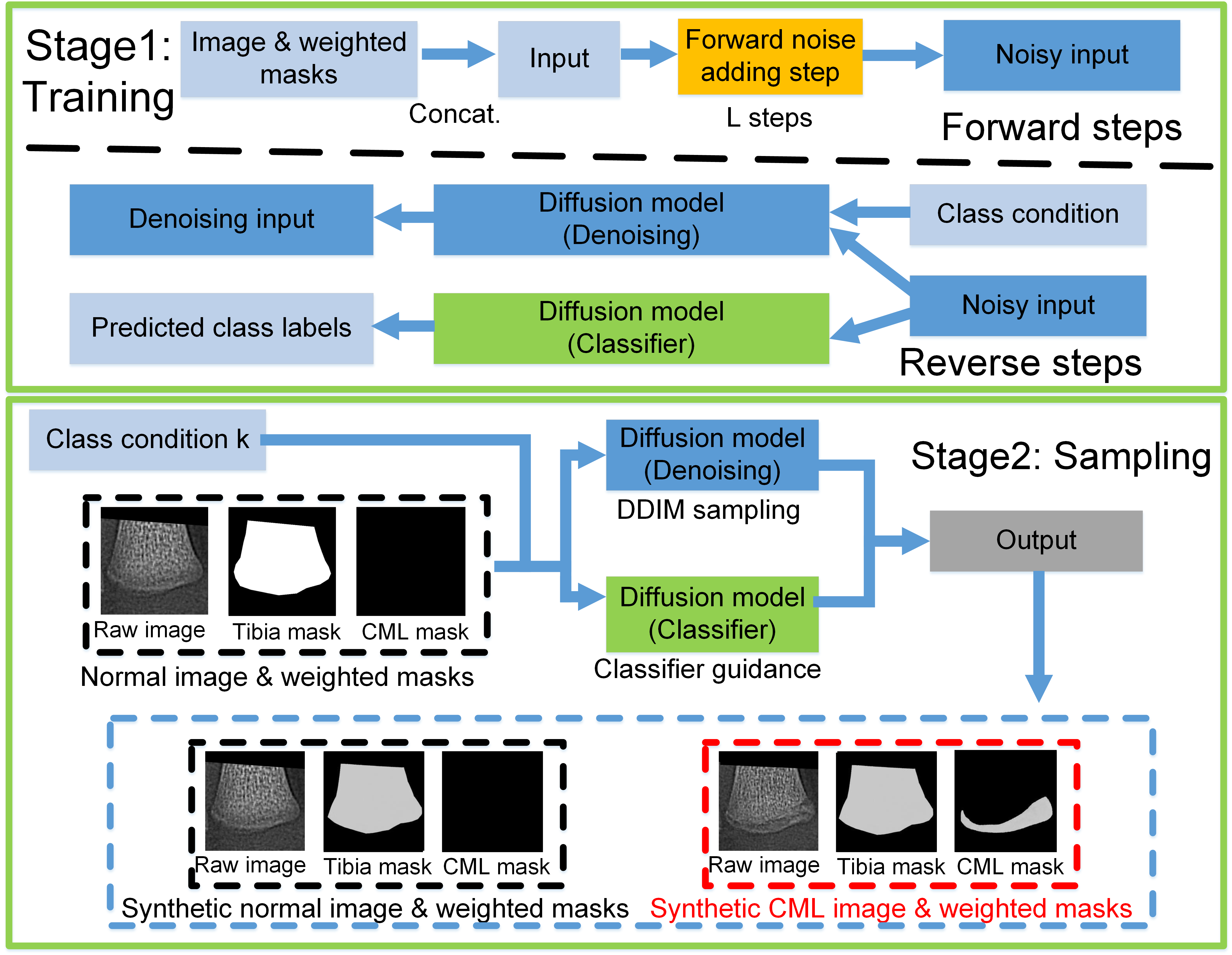}
\caption{Algorithmic structure of the MaC-DM, which consists of two stages. In the first stage, noise is gradually added to both training images and masks (forward process). Two separate networks are then trained: one for classification and another for denoising (reverse process). In the second stage, a new synthetic image and its associated masks are sampled from MaC-DM based on the input image from the real dataset and by incorporating mask-based condition along with classifier guidance.} \label{fig1}
\end{figure}

\subsection{Image generation via conditional diffusion model}
\subsubsection{Classifier guidance} It is desirable for a diffusion model to generate synthetic images with specific attributes such as explicit class labels or with certain properties. This can be accomplished by injecting conditional information into the diffusion model. To enable this attractive feature, Dhariwal et al. \cite{dhariwal2021diffusion} trained a classifier using noisy images and used the gradient of the classifier to guide the diffusion model. In particular, given the conditional label $y$ for a diffusion model $\epsilon_\theta$, they constructed a modified score function by integrating %$p(\mathbf{x}_t)p(y\vert\mathbf{x}_t)$, 
$p(y\vert\mathbf{x}_t)$, the posterior distribution for classification. This modified score function is given by
\begin{equation}
\nabla_{x_t}\log (p_\theta(\mathbf{x}_t)p_\phi (y\vert\mathbf{x}_t))=\nabla_{x_t}\log p_\theta(\mathbf{x}_t)+\nabla_{x_t}\log p_\phi (y\vert\mathbf{x}_t)
\end{equation}
where $\nabla_{x_t}\log p_\theta(\mathbf{x}_t)=-\frac{1}{\sqrt{1-\bar{\alpha}_{t}}}\epsilon_\theta({x}_t)$. Then, the predicted noise obtained from classifier-guided model becomes 
$
\bar\epsilon_\theta({x}_t)=\epsilon_\theta({x}_t)-\sqrt{1-\bar{\alpha}_{t}}g\nabla_{x_t}\log p_\phi (y\vert\mathbf{x}_t)
$, where $\nabla_{x_t}\log p_\phi (y\vert\mathbf{x}_t)$ is the gradient of the classifier, and $g$ is a weighting factor to adjust the relative strengths of the gradient for classifier guidance.

\subsubsection{Masked conditional classifier guidance} Training a conditional diffusion model with simple class labels (e.g. CML versus normal) may not be sufficient to characterize the geometry of the distal tibia, such as the bone shape and the CML fracture pattern. This stems from the large shape variation of the distal tibia and the CML fracture. To address this deficiency, we explicitly introduce the bone shape and fracture region into the diffusion model by injecting two different segmentation masks as conditions. Motivated by previous work \cite{rombach2022high}, we introduced the image conditions by channel-wise concatenation of the conditional images. Specifically, we derive the input of the diffusion model as
$I_{t} =concat(w_{1}\mathbf{x}_t,w_{2}\mathbf{b}_t,w_{3}\mathbf{c}_t)$
where $\mathbf{x}_t$, $\mathbf{b}_t$, and $\mathbf{c}_t$ are the noisy real image, the tibia segmentation mask, and the CML fracture segmentation mask at time step $t$, respectively; and $w_{1}$, $w_{2}$, $w_{3}$ are the weighting factors that balance the relative contributions of the various input channels. To capture the prominent features of the CML for data generation, we trained another diffusion model with mask conditions as input for classifier guidance. By substituting $I_{t}$ into equation (4), the new score function of our mask conditional diffusion model with classifier guidance is given by
$
\nabla_{I_t}\log (p_\theta({I}_t)p_\phi (y\vert{I}_t))=\nabla_{I_t}\log p_\theta({I}_t)+\nabla_{I_t}\log p_\phi (y\vert{I}_t) .
$
With the two different segmentation masks as conditions to the diffusion model, the overall loss function of our proposed method becomes:
\begin{equation}
L_\text{total}= \mathbb{E}_{\epsilon \sim \mathcal{N}(0,\mathbf I)}[\| \epsilon-\epsilon_\theta(w_{1}x_{t},w_{2}b_t, w_{3}c_t,t)\|^{2}_2]
\end{equation}
where $\epsilon_\theta$ is the trained diffusion model. To improve the sampling speed, we adopted a hybrid loss objective for training with the weighted sum of $L_\text{total}+\lambda L_\text{vlb}$ in \cite{nichol2021improved}. We set a relatively smaller scaling weights to the two segmentation masks with $w_{1}=1.0$, $w_{2}=0.8$ and $w_{3}=0.8$. Empirically, we found that larger weights of the segmentation masks reduced the image reconstruction quality.

\subsubsection{Image-to-image translation for data generation} After training a classifier and a denoising diffusion model with mask conditions, we utilized the proposed MaC-DM for image-to-image translation, when given a class condition $K$, where $K \in \{0,1\}$ (i.e., normal or CML). As there are disproportionately more normal radiographs than CML ones in our training datasets, we only used the normal radiographs of the distal tibia as inputs for image translation (i.e. we translated normal-to-CML and normal-to-normal radiographs). This approach generated more augmented CML images to address the class imbalance problem for binary classification. To preserve the overall structure of the tibia during the denoising process, we only used a noisy image $x_{t}$ from an intermediate time step $Z$, where $Z<T$ for image generation. 

\section{Experiments}
\subsubsection{Datasets} We curated two different datasets of the distal tibial radiograph over two distinct time periods. The primary dataset consisted of 178 normal radiographs and 74 with CMLs ${(2009}$--${2021)}$. The secondary dataset consisted of 45 normal radiographs and 8 with CMLs ${(2006}$--${2008}$ and ${2021}$--${2022)}$. Given that these radiographs were obtained over such a long period of time, different X-ray imaging techniques and imaging systems were utilized, resulting in varying image contrast and resolution. This existential image heterogeneity is expected, and underscores the importance of having an algorithm that generalizes to radiographs of differing image qualities. The original 1024${\times}$1024 gray-scale radiographs were cropped to the distal one-fifth of the tibia to focus over the region of concern (see S.1 in Supplementary Material for details). The resultant images were resized to 256${\times}$256 pixels, and normalized to an intensity range of 0 to 1. A radiologist vetted these radiographs in establishing the ground truth (i.e. normal versus CML). Additionally, for each radiograph, the distal tibia and the CML fracture were manually segmented by a radiologist in establishing the ground truth.

\subsubsection{Implementation} We trained the DDPM using an improved version of U-Net architecture \cite{nichol2021improved}. Specifically, we modified this network to have a three-channel input (gray-scale radiograph, and the weighted distal tibial and CML fracture segmentation masks). We trained the conditional diffusion model with a batch size of 5 and 120,000 iterations; and trained the classifier with a batch size of 10 and 150,000 iterations. All the DDPMs were trained using an Adam optimizer, a learning rate of $10^{-4}$, and a sampling time step $t$ = 1000. The training process took approximately 1.5 days using a Quadro RTX 8000 GPU. All networks were implemented using Pytorch version 1.7.1.

To evaluate the added-value of MaC-DM, we compared the accuracy of five ResNet-34 \cite{he2016deep} binary classifiers. We conducted a 5-fold cross validation (CV) experiment to evaluate each classifier. The first classifier was trained with the primary dataset using standard data augmentation (random cropping). To compare our method with other state-of-the-art (SOTA) methods for image-to-image translation, the rest of the four classifiers were trained with augmented data (N = 472, with 188 real and 284 augmented synthetic data) generated from CycleGAN \cite{zhu2017unpaired}, AttentionGAN \cite{tang2021attentiongan}, the baseline diffusion models without mask conditions \cite{dhariwal2021diffusion}, and our proposed MaC-DM. For MaC-DM and baseline diffusion models, we utilized the DDIM to generate synthetic CML and normal radiographs using normal real radiographs from the primary training dataset as input. To evaluate the generalizability of these classifiers, we employed the trained ResNet that achieved the highest validation performance during the 5-fold CV experiment, and applied it to the secondary dataset (which has a different distribution than the primary dataset used in the 5-fold CV). 

We evaluated the utility of the CML segmentation mask generated by MaC-DM in two manners. First, we leveraged MaC-DM to generate synthetic CML radiographs and their corresponding tibial and CML segmentation masks, using 80\% of radiographs from the primary dataset as input. We then trained a U-Net model \cite{ronneberger2015u} for CML segmentation using a combination of real and synthetic CML images/masks. We evaluated the segmentation performance using the remaining 20\% of the primary data for testing. Second, to objectively evaluate the image quality in terms of fidelity and diversity \cite{dhariwal2021diffusion}, we computed the Fréchet inception distance (FID) \cite{heusel2017gans} to compare the data distribution between the real images (primary dataset) and the synthetic images generated from normal-to-CML translation task in the 5-fold CV.
\section{Results and Discussion} 
\begin{figure}[t!]
\includegraphics[width=\textwidth]{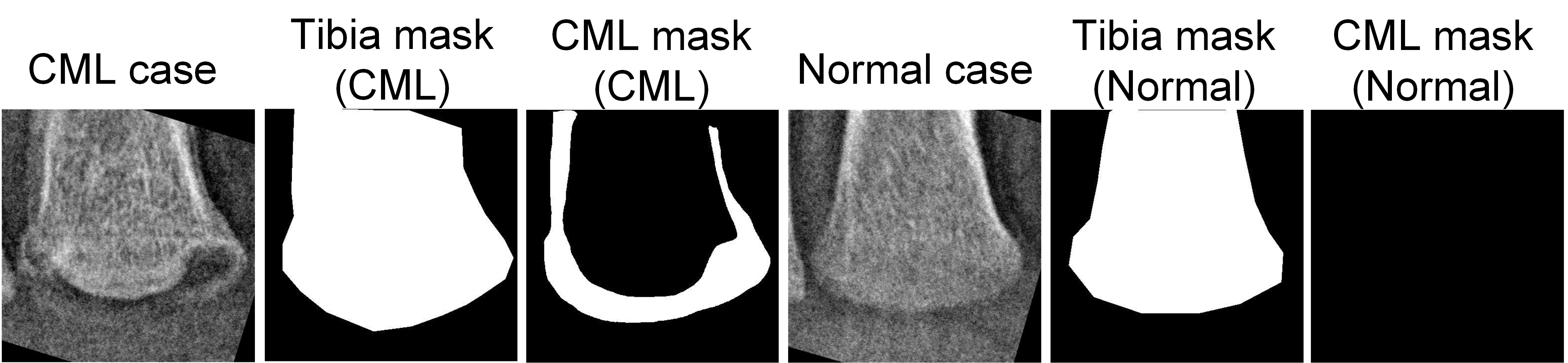}
\caption{The distal tibial CML image and its associated tibial and CML segmentation masks are shown in the first three frames. The normal distal tibial image and its associated tibial and CML segmentation masks are shown in the last three frames.} \label{fig2}
\end{figure}

The weighting factors related to the segmentation masks of tibia $w_{2}$ and CML fractures $w_{3}$, as defined in equation (5), are important hyperparameters that affect the quality of the synthetic images. To optimize these, we initially set the noise level $Z$ to 600 and the gradient scale for classifier guidance $g$ to 100. Next, we trained our MaC-DM on the primary dataset using various combinations of $w_{2}$ and $w_{3}$, ranging from 0.2 to 1.0. We then applied the normal-to-CML image translation task to all normal images in the secondary dataset. A blinded radiologist graded the image quality of the generated images for each combination of weights, assigning a score of 1 for excellent image quality and 0 otherwise. In this parametric study, $w_{2}$=0.8 and $w_{3}$=0.8 received the highest summation scores. Using a similar approach, we identified the optimal noise level of $Z$=800 and the scale of $g$=300 for classifier guidance.

\begin{figure}[t]
\includegraphics[width=\textwidth]{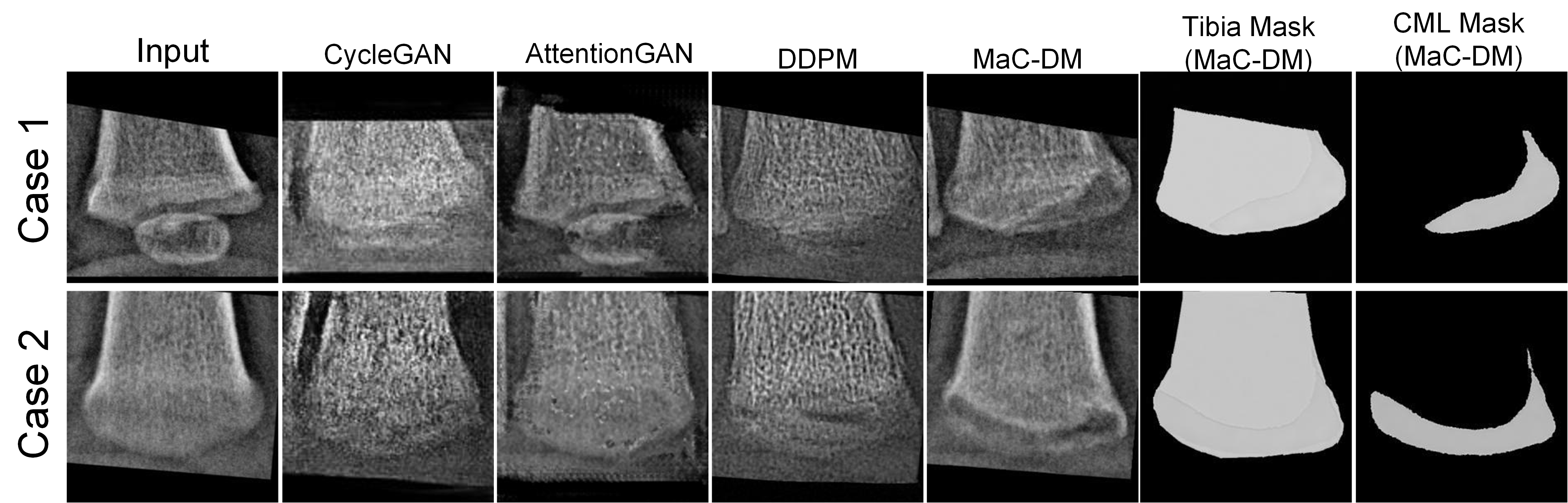}
\caption{Example images and segmentation masks generated by MaC-DM, in comparison to images from other  generative models. A radiologist checked the quality of the images and concluded that MaC-DM generated more realistic and higher quality CML images than other generative models. In addition, MaC-DM generated tibial and CML segmentation masks without any additional costs.} \label{fig3}
\end{figure}

Based on our 5-fold CV experiment, our proposed data augmentation method outperformed the other four methods in classifying CML radiographs from normal ones, in terms of accuracy, sensitivity, and specificity (Table~\ref{tab1}). When trained on the primary dataset, the proposed MaC-DM achieved the highest accuracy and sensitivity when tested on the secondary dataset, indicating improved generalizability. Of note, although the baseline DDPM method had a slightly better specificity on the secondary dataset than ours (93.3\% versus 91.1\%), its sensitivity (62.5\% versus 87.5\%) was much lower. Fig.~\ref{fig3} shows that the images produced by other generative methods have blurry bony margins and often contained artifacts, whereas the augmented images from MaC-DM showed uncanny realism.

Table~\ref{tab2} (top) summarized the CML segmentation results via U-Net. The augmented images and their corresponding synthetic segmentation masks generated by MaC-DM improved the segmentation performance of U-Net. Surprisingly, even when training solely on the synthetic images and their generated segmentation masks, MaC-DM was able to improve the U-Net's segmentation performance. These findings suggest that the synthetic images and their segmentation labels produced by MaC-DM are valuable in enhancing the performance of CML segmentation, at no additional labeling cost. In the case of normal-to-CML image translation, MaC-DM achieved an FID of 89.87 (Table~\ref{tab2} (bottom)), outperforming other SOTA methods. Our small training dataset likely hindered other SOTA methods from generating quality synthetic images as it typically requires large datasets for training. These results highlight the importance of our method in synthesizing realistic CML images from a small training dataset.

\begin{table}
\caption{Performance summary of the five classifiers in terms of accuracy, sensitivity, and specificity. The 5-fold CV experiment was conducted using the primary dataset, while independent testing was conducted using the secondary dataset.}\label{tab1}
\begin{tabular}{|l|lllllll|}
\hline
            & \multicolumn{7}{c|}{\it 5-fold cross validation}                                                                                                                                                                                                                                                                         \\ \hline
Method      & \multicolumn{1}{l|}{ResNet}  & \multicolumn{2}{l|}{\begin{tabular}[c]{@{}l@{}}ResNet + \\ CycleGAN\cite{zhu2017unpaired}\end{tabular}} & \multicolumn{2}{l|}{\begin{tabular}[c]{@{}l@{}}ResNet +\\ AttentionGAN\cite{tang2021attentiongan}\end{tabular}}  & \multicolumn{1}{l|}{\begin{tabular}[c]{@{}l@{}}ResNet +\\  DDPM w/o mask\cite{dhariwal2021diffusion}\end{tabular}} & Ours             \\ \hline
Accuracy    & \multicolumn{1}{l|}{93.00\%} & \multicolumn{2}{l|}{95.20\%}                                                      & \multicolumn{2}{l|}{94.00\%}                                                          & \multicolumn{1}{l|}{96.00\%}                                                           & \textbf{96.40\%} \\ \hline
Sensitivity & \multicolumn{1}{l|}{88.00\%} & \multicolumn{2}{l|}{90.71\%}                                                      & \multicolumn{2}{l|}{90.36\%}                                                          & \multicolumn{1}{l|}{88.90\%}                                                           & \textbf{90.73\%} \\ \hline
Specificity & \multicolumn{1}{l|}{96.00\%} & \multicolumn{2}{l|}{97.11\%}                                                      & \multicolumn{2}{l|}{95.48\%}                                                          & \multicolumn{1}{l|}{98.84\%}                                                           & \textbf{98.84\%} \\ \hline \hline
            & \multicolumn{7}{c|}{\textit{Independent testing}}                                                                                                                                                                                                                                                                    \\ \hline
Method      & \multicolumn{1}{l|}{ResNet}  & \multicolumn{2}{l|}{\begin{tabular}[c]{@{}l@{}}ResNet + \\ CycleGAN\cite{zhu2017unpaired}\end{tabular}} & \multicolumn{2}{l|}{\begin{tabular}[c]{@{}l@{}}ResNet +\\  AttentionGAN\cite{tang2021attentiongan}\end{tabular}} & \multicolumn{1}{l|}{\begin{tabular}[c]{@{}l@{}}ResNet + \\ DDPM w/o mask\cite{dhariwal2021diffusion}\end{tabular}} & Ours             \\ \hline
Accuracy    & \multicolumn{1}{l|}{84.90\%} & \multicolumn{2}{l|}{84.90\%}                                                      & \multicolumn{2}{l|}{86.80\%}                                                          & \multicolumn{1}{l|}{88.67\%}                                                           & \textbf{90.60\%} \\ \hline
Sensitivity & \multicolumn{1}{l|}{75.00\%} & \multicolumn{2}{l|}{62.5\%}                                                       & \multicolumn{2}{l|}{62.50\%}                                                          & \multicolumn{1}{l|}{62.50\%}                                                           & \textbf{87.50\%} \\ \hline
Specificity & \multicolumn{1}{l|}{86.67\%} & \multicolumn{2}{l|}{88.90\%}                                                      & \multicolumn{2}{l|}{91.10\%}                                                          & \multicolumn{1}{l|}{\textbf{93.33\%}}                                                  & 91.10\%          \\ \hline
\end{tabular}
\end{table}
\begin{table}[h!]
\caption{Performance summary of the segmentation results based on mean dice score (top). The image quality evaluation results were based on FID score (bottom).}\label{tab2}
\begin{tabular}{|l|lllllllllllllll|}
\hline
                 & \multicolumn{15}{c|}{\it Segmentation evaluation}                                                                                                 \\ \hline
Method (U-Net)   & \multicolumn{5}{l|}{Real images} & \multicolumn{5}{l|}{Augmented images}                 & \multicolumn{5}{l|}{Real + augmented images}         \\ \hline
Dice coefficient & \multicolumn{5}{l|}{0.74±0.03}       & \multicolumn{5}{l|}{0.83±0.05}                             & \multicolumn{5}{l|}{\textbf{0.85±0.04}}                  \\ \hline \hline
                 & \multicolumn{15}{c|}{\it FID score evaluation (lower is better)}                                                                                                \\ \hline
Method           & \multicolumn{6}{l|}{CycleGAN\cite{zhu2017unpaired}}           & \multicolumn{3}{l|}{AttentionGAN\cite{tang2021attentiongan}} & \multicolumn{3}{l|}{DDPM w/o mask\cite{dhariwal2021diffusion}} & \multicolumn{3}{l|}{Ours}  \\ \hline
FID score        & \multicolumn{6}{l|}{165.33}             & \multicolumn{3}{l|}{139.30}       & \multicolumn{3}{l|}{99.71}         & \multicolumn{3}{l|}{\textbf{89.87}} \\ \hline
\end{tabular}
\end{table}

\section{Conclusions} We proposed a novel data augmentation method for CML fracture classification and segmentation. Our proposed method overcame the limitations of previous techniques in generating diverse radiographic images of distal tibial CML and produced realistic synthetic images with associated segmentation labels. This method has the potential to improve the accuracy and robustness of machine learning models for the diagnosis of infant abuse.

\subsubsection{Acknowledgements} This work was supported in part by the Society for Pediatric Radiology Research and Education Foundation Pilot Grant, National Institute of Child Health and Human Development (No. R21HD108634), National Institute of Diabetic and Digestive and Kidney Diseases (No. R21DK123569 and R01DK125561), and National Institute of Biomedical Imaging and Bioengineering (No. R21EB029627). 

%
% ---- Bibliography ----
%
% BibTeX users should specify bibliography style 'splncs04'.
% References will then be sorted and formatted in the correct style.
%

%
\bibliographystyle{splncs04}
\bibliography{sample}

\begin{thebibliography}{10}
\providecommand{\url}[1]{\texttt{#1}}
\providecommand{\urlprefix}{URL }
\providecommand{\doi}[1]{https://doi.org/#1}

\bibitem{coley2013caffey}
Coley, B.D.: Caffey's pediatric diagnostic imaging e-book. Elsevier Health
  Sciences (2013)

\bibitem{dhariwal2021diffusion}
Dhariwal, P., Nichol, A.: Diffusion models beat gans on image synthesis.
  Advances in Neural Information Processing Systems  \textbf{34},  8780--8794
  (2021)

\bibitem{flaherty2014evaluating}
Flaherty, E.G., Perez-Rossello, J.M., Levine, M.A., Hennrikus, W.L.,
  of~Pediatrics Committee~on Child~Abuse, A.A., Neglect, on~Radiology, S.,
  ENDOCRINOLOGY, S.O., ORTHOPAEDICS, S.O., for Pediatric~Radiology, S.,
  Christian, C.W., et~al.: Evaluating children with fractures for child
  physical abuse. Pediatrics  \textbf{133}(2),  e477--e489 (2014)

\bibitem{goodfellow2020generative}
Goodfellow, I., Pouget-Abadie, J., Mirza, M., Xu, B., Warde-Farley, D., Ozair,
  S., Courville, A., Bengio, Y.: Generative adversarial networks.
  Communications of the ACM  \textbf{63}(11),  139--144 (2020)

\bibitem{han2021madgan}
Han, C., Rundo, L., Murao, K., Noguchi, T., Shimahara, Y., Milacski, Z.{\'A}.,
  Koshino, S., Sala, E., Nakayama, H., Satoh, S.: Madgan: Unsupervised medical
  anomaly detection gan using multiple adjacent brain mri slice reconstruction.
  BMC bioinformatics  \textbf{22}(2),  1--20 (2021)

\bibitem{he2016deep}
He, K., Zhang, X., Ren, S., Sun, J.: Deep residual learning for image
  recognition. In: Proceedings of the IEEE conference on computer vision and
  pattern recognition. pp. 770--778 (2016)

\bibitem{heusel2017gans}
Heusel, M., Ramsauer, H., Unterthiner, T., Nessler, B., Hochreiter, S.: Gans
  trained by a two time-scale update rule converge to a local nash equilibrium.
  Advances in neural information processing systems  \textbf{30} (2017)

\bibitem{ho2020denoising}
Ho, J., Jain, A., Abbeel, P.: Denoising diffusion probabilistic models.
  Advances in Neural Information Processing Systems  \textbf{33},  6840--6851
  (2020)

\bibitem{kearney2020attention}
Kearney, V., Ziemer, B.P., Perry, A., Wang, T., Chan, J.W., Ma, L., Morin, O.,
  Yom, S.S., Solberg, T.D.: Attention-aware discrimination for mr-to-ct image
  translation using cycle-consistent generative adversarial networks.
  Radiology: Artificial Intelligence  \textbf{2}(2),  e190027 (2020)

\bibitem{kim2022diffusion}
Kim, B., Ye, J.C.: Diffusion deformable model for 4d temporal medical image
  generation. In: Medical Image Computing and Computer Assisted
  Intervention--MICCAI 2022: 25th International Conference, Singapore,
  September 18--22, 2022, Proceedings, Part I. pp. 539--548. Springer (2022)

\bibitem{kleinman1986metaphyseal}
Kleinman, P.K., Marks, S., Blackbourne, B.: The metaphyseal lesion in abused
  infants: a radiologic-histopathologic study. American Journal of
  Roentgenology  \textbf{146}(5),  895--905 (1986)

\bibitem{kleinman1995relationship}
Kleinman, P.K., Marks~Jr, S.C.: Relationship of the subperiosteal bone collar
  to metaphyseal lesions in abused infants. JBJS  \textbf{77}(10),  1471--1476
  (1995)

\bibitem{kleinman1995inflicted}
Kleinman, P.K., Marks~Jr, S.C., Richmond, J.M., Blackbourne, B.D.: Inflicted
  skeletal injury: a postmortem radiologic-histopathologic study in 31 infants.
  AJR. American journal of roentgenology  \textbf{165}(3),  647--650 (1995)

\bibitem{kleinman2011prevalence}
Kleinman, P.K., Perez-Rossello, J.M., Newton, A.W., Feldman, H.A., Kleinman,
  P.L.: Prevalence of the classic metaphyseal lesion in infants at low versus
  high risk for abuse. American Journal of Roentgenology  \textbf{197}(4),
  1005--1008 (2011)

\bibitem{ChildMaltreatment}
Maltreatment, C.: Children’s Bureau, Administration on Children, Youth, and
  Family (2018)

\bibitem{nichol2021improved}
Nichol, A.Q., Dhariwal, P.: Improved denoising diffusion probabilistic models.
  In: International Conference on Machine Learning. pp. 8162--8171. PMLR (2021)

\bibitem{peng2022towards}
Peng, C., Guo, P., Zhou, S.K., Patel, V.M., Chellappa, R.: Towards performant
  and reliable undersampled mr reconstruction via diffusion model sampling. In:
  Medical Image Computing and Computer Assisted Intervention--MICCAI 2022: 25th
  International Conference, Singapore, September 18--22, 2022, Proceedings,
  Part VI. pp. 623--633. Springer (2022)

\bibitem{rombach2022high}
Rombach, R., Blattmann, A., Lorenz, D., Esser, P., Ommer, B.: High-resolution
  image synthesis with latent diffusion models. In: Proceedings of the IEEE/CVF
  Conference on Computer Vision and Pattern Recognition. pp. 10684--10695
  (2022)

\bibitem{ronneberger2015u}
Ronneberger, O., Fischer, P., Brox, T.: U-net: Convolutional networks for
  biomedical image segmentation. In: Medical Image Computing and
  Computer-Assisted Intervention--MICCAI 2015: 18th International Conference,
  Munich, Germany, October 5-9, 2015, Proceedings, Part III 18. pp. 234--241.
  Springer (2015)

\bibitem{saxena2021generative}
Saxena, D., Cao, J.: Generative adversarial networks (gans) challenges,
  solutions, and future directions. ACM Computing Surveys (CSUR)
  \textbf{54}(3),  1--42 (2021)

\bibitem{servaes2016etiology}
Servaes, S., Brown, S.D., Choudhary, A.K., Christian, C.W., Done, S.L., Hayes,
  L.L., Levine, M.A., Moreno, J.A., Palusci, V.J., Shore, R.M., et~al.: The
  etiology and significance of fractures in infants and young children: a
  critical multidisciplinary review. Pediatric radiology  \textbf{46},
  591--600 (2016)

\bibitem{song2020denoising}
Song, J., Meng, C., Ermon, S.: Denoising diffusion implicit models. arXiv
  preprint arXiv:2010.02502  (2020)

\bibitem{tang2021attentiongan}
Tang, H., Liu, H., Xu, D., Torr, P.H., Sebe, N.: Attentiongan: Unpaired
  image-to-image translation using attention-guided generative adversarial
  networks. IEEE transactions on neural networks and learning systems  (2021)

\bibitem{tsai2014high}
Tsai, A., McDonald, A.G., Rosenberg, A.E., Gupta, R., Kleinman, P.K.:
  High-resolution ct with histopathological correlates of the classic
  metaphyseal lesion of infant abuse. Pediatric radiology  \textbf{44},
  124--140 (2014)

\bibitem{wolleb2022diffusion}
Wolleb, J., Bieder, F., Sandk{\"u}hler, R., Cattin, P.C.: Diffusion models for
  medical anomaly detection. In: Medical Image Computing and Computer Assisted
  Intervention--MICCAI 2022: 25th International Conference, Singapore,
  September 18--22, 2022, Proceedings, Part VIII. pp. 35--45. Springer (2022)

\bibitem{zhu2017unpaired}
Zhu, J.Y., Park, T., Isola, P., Efros, A.A.: Unpaired image-to-image
  translation using cycle-consistent adversarial networks. In: Proceedings of
  the IEEE international conference on computer vision. pp. 2223--2232 (2017)

\end{thebibliography}

\end{document}